# XAI FL-IDS: A Federated Learning and SHAP-Based Explainable Framework for Distributed Intrusion Detection Systems

Mohammad Hossein Gholamrezazadeh*
Faculty of Computer Engineering
University of Isfahan
Isfahan, Iran
gholamrezazadehmohammad@eng.ui.ac.ir

AhmadReza Montazerolghaem
Faculty of Computer Engineering
University of Isfahan
Isfahan, Iran
a.montazerolghaem@comp.ui.ac.ir

***Abstract*: An Intrusion Detection System (IDS) is vital in cybersecurity, detecting unauthorized activity across networks. With attacks on network layers increasing, stronger IDSs are needed. Yet most IDSs rely on centralized detection, forcing IoT nodes to ship data to a server—adding overhead and offering no privacy guarantees. Moreover, conventional models focus solely on flagging attacks, without explaining how individual features influence those decisions. This research aims to address these dual limitations by first proposing a solution for privacy preservation and then adding explainability to the new system. We introduce an innovative framework called XAI FL-IDS, which integrates Federated Learning (FL) with Explainable AI (XAI). The XAI FL-IDS system eliminates concerns over data transfer because each node trains its data locally and only sends the necessary update parameters to the server. Additionally, all detections, both at the local node and central server levels, are scrutinized using SHapley Additive exPlanations (SHAP), providing detailed insight into the decision-making process. This system consists of a central server and 10 clients and utilizes the Edge-IIoTset dataset, which is distributed among all clients with careful attention paid to class balancing. On each client, the XGBoost model is executed on local data. The proposed method demonstrates robust efficiency and strong performance in intrusion detection, achieving an accuracy of over 99% and, at times, reaching 100%. By incorporating FL, the confidentiality of the network information on every local node is guaranteed.**

## I. INTRODUCTION

Following the rapid advancement of communication technologies, IoT devices have shifted from isolation to becoming interconnected with the internet. According to recent reports, the number of devices connected to the internet is projected to reach 70 billion by the year 2025. This growth, fueled by the Fourth Industrial Revolution (Industry 4.0) and the penetration of IoT into various sectors, has led to a significant research focus on integrating Artificial Intelligence (AI) and Big Data technologies with IoT infrastructure. The sheer volume of data generated and collected across these IoT networks constitutes a true Big Data challenge. The core objective of gathering data from different industrial levels is to extract useful knowledge; however, the quality and security of IoT services fundamentally rely on data integrity and accuracy. Unfortunately, this integrity can be severely compromised by the injection of malicious events—such as incorrect data and various attacks—or by network-level failures.

Given the explanations presented, the critical necessity for robust IDS is highly palpable. One of the most effective implementation approaches involves intelligent methods, specifically those utilizing Machine Learning (ML) and Deep Learning (DL) to perform this essential detection.

A significant challenge faced by these intelligent detection systems is privacy preservation. Conventionally, each IoT network node must transmit its raw data to a central server for the model to be trained on the aggregate data. This transfer not only presents a security risk, compromising the privacy of individual nodes, but also imposes considerable overhead on the system and the network. To overcome these limitations, this research utilizes FL, eliminating the need to transmit raw data to the central server by performing training locally on each node. Crucially, the FL approach in this paper incorporates a novel modification: during the aggregation phase, the system identifies the best model from the previous round and broadcasts only that superior model to all nodes for the next iteration.

Another limitation inherent in all IDS that utilize intelligent methods is the black box nature of the underlying ML and DL approaches. By black box, it is meant that these models can only identify the type of attack without providing insight into the decision-making process. However, contemporary requirements necessitate knowing why a particular detection was made, as understanding the cause can significantly impact system performance and utility across various applications. This research aims to resolve this issue and provide an explainable system by incorporating SHAP.

## II. RELATED WORK

A study by Albanbay et al. [7] evaluated three distinct architectures—Deep Neural Network (DNN), Convolutional Neural Network (CNN), and CNN + Bidirectional Long Short-Term Memory (BiLSTM)— for Intrusion Detection in IoT environments. They concluded that the hybrid CNN + BiLSTM model achieved the highest accuracy (up to 99%), owing to its ability to extract both spatial features (via CNN) and temporal dependencies (via BiLSTM) from the network traffic. However, the CNN alone demonstrated the best balance between performance and computational efficiency with approximately 98% accuracy, making it more suitable for resource-constrained devices, such as the Raspberry Pi 5.

Rashid et al. [8] utilized CNN and Recurrent Neural Network (RNN) models for Intrusion Detection in IIoT networks, achieving an accuracy of 92.49%. This figure was closely aligned with the performance of a centralized system (93.92%) while simultaneously offering the crucial advantage of privacy preservation. Furthermore, in the Transportation IoT sector, Bhavsar et al. [6] introduced a modular Federated Learning-Based Intrusion Detection System (FL-IDS) framework. This system employed a combination of Logistic Regression (LR) and a Pearson Correlation Coefficient-Convolutional Neural Network (PCC-CNN), which yielded an exceptionally high accuracy of 99.93% on the Car-Hacking dataset.

A newer approach involves combining FL with Optimization Algorithms to enhance accuracy. Karunamurthy et al. [9] utilized the Chimp Optimization Algorithm (COA) for optimal feature selection. This was done to reduce the dimensionality of the network traffic data and improve the classification accuracy of the local CNN to 95.59%.

Thein et al. [1] introduced a Personalized FL-IDS utilizing a CNN and a Mini-batch Logit Adjustment Loss. This methodology yields a personalized model tailored to the local data distribution of each client, thereby improving the performance of the central model under heterogeneous conditions, as it avoids relying on a single fixed model across varied data distributions.

Alsaleh et al. [3] presented a different strategy to handle both system and data heterogeneity: a Semi-Decentralized FL model. This model reduces communication overhead by clustering clients using Autoencoder (AE)-based dimensionality reduction and then delegating the aggregation task to the head node of each cluster. By employing Bi-LSTM as the local model, this architecture demonstrated high performance in identifying DDoS attacks across several heterogeneous IoT datasets.

In addition to managing Non-IID data, Thein et al. [1] proposed a server-side malicious client detection mechanism that uses the cosine similarity of local models to identify and prevent poisoning agents from being aggregated into the central model. This is achieved through a two-stage process: First, local models are compared with the pre-computed global model to identify potentially benign clients. Second, their angular deviation from the centroid of the non-malicious clients is measured. Clients exhibiting low cosine similarity are identified as compromised and are excluded from the global model aggregation process, thereby neutralizing poisoning attacks without degrading performance.

Gutti et al. [4] specifically focused on resilience against adversarial attacks and privacy preservation. They introduced a novel hybrid adaptive-weight aggregation algorithm called HADA (Hybrid Adaptive-Weight Aggregation). HADA weights the client updates based on two factors: the SHAP-based feature stability and the client's Individual Differential Privacy (DP) budget.

Fatema et al. [5] introduced an explicit FL-XAI framework named FEDXAIIDS, which integrates FL with SHAP. This model utilized an Artificial Neural Network (ANN) as the local model and applied SHAP to the aggregated output to identify the contribution of network features to attack detection. This analysis revealed that the UDP feature had the most significant impact on detection. While this approach demonstrated lower accuracy (approximately 88.2%) compared to some centralized models, it successfully emphasized the critical importance of interpretability alongside privacy preservation.

Gutti et al. [4] also employed SHAP, but their objective was dual: First, to use SHAP in the pre-processing stage for stable feature selection to enhance the model's resilience against adversarial attacks; and second, to use it at the core of the HADA aggregation algorithm for client weighting. This mechanism ensures the transparency of the IDS while maintaining security and scalability.

Finally, all the items reviewed in the related work section are presented in Table 1. This table examines aspects such as the method and algorithms (which are the same as the AI methods) used in the articles, the dataset utilized in each study, and the accuracy and performance of the proposed systems.

## III. PROPOSED METHODS

### *A. Dataset Overview*

Edge-IIoTset [1] is a comprehensive, realistic cybersecurity dataset for IoT/IIoT, designed for ML-based intrusion detection in both centralized and federated modes. Built on a seven-layer testbed (Cloud, NFV, Blockchain, Fog, SDN, Edge, and IoT/IIoT Perception), it gathers data from over 10 IoT device types and features 14 attacks across five threat categories: DoS/DDoS, information gathering, man-in-the-middle, injection, and malware. The dataset contains 63 columns and 157,800 rows.

### *B. Data Processing*

The first stage focused on handling missing values. NaN replaced the infinite values within the dataset, and irrelevant or incomplete features were simultaneously removed. Among the features in this dataset, a subset lacked numerical values and was defined as categorical. In this step, the categorical data are converted to numerical data, all NaN values are imputed by substituting them with the Mode of their respective class or category. Following these transformations, all features were normalized, and the data was ready for splitting.

After the 80/20 train/test split, the training data are distributed among clients, with careful attention paid to class balancing during this process. Subsequently, each client prepared its respective portion of the data, performing its own internal splitting for local model training.

### *C. Federated Learning*

FL has recently emerged as a promising approach to tackle the dual challenges of data privacy and scalability in distributed environments.

[1] Edge-IIoTset Cyber Security Dataset of IoT & IIoT

*Table 1: Comparative Summary of Methods, Data, and Performance in Related FL-IDS Work*

| **Authors, Year, and Article Title** | **Method and Algorithm** | **Datasets** | **Accuracy and Performance** |
|---|---|---|---|
| Anwer et al. (2025)[2]: Advanced Intrusion Detection in the IIoT using FL and LSTM models | Hybrid CNN-Bi-LSTM + FL (FedAvg) | X-IIoTID, WUSTL-IIoT, Edge-IIoTset | 97.8% Accuracy on X-IIoTID; AUC exceeding 95% |
| Albanbay et al. (2025)[7]: FL-Based IDS in IoT networks: Performance evaluation and data scaling study | DNN, CNN, Hybrid CNN + BiLSTM + FL | CICIoT2023 | Approx. 98% Accuracy (CNN); Approx. 99% Accuracy (CNN + BiLSTM) |
| Rashid et al. (2023)[8]: A FL-based approach for improving IDS in IIoT networks | FL (FedAvg) + CNN and RNN | Edge-IIoTset | 92.49% Accuracy (with RNN) |
| Bhavsar et al. (2024)[6]: FL-IDS using edge devices for transportation IoT | FL-IDS (LR , PCC-CNN) | NSL-KDD, Car-Hacking | Overall Accuracy: 94% - 99%; 99.93% (PCC-CNN on Car-Hacking) |
| Karunamurthy et al. (2025)[9]: An optimal FL-based IDS for IoT environment | FL-IDS (CNN) + Chimp Optimization Algorithm (COA) | MQTT dataset | Maximum Detection Accuracy of 95.59% |
| Thin Tharaphe Thein et al. (2024)[1]: Personalized federated learning-based intrusion detection system: Poisoning attack and defense | pFL-IDS (FL): 1D-CNN + logit adjustment; server-side poisoned-client detector | N-BaIoT (mini N-BaIoT constructed from original dataset). | No-attack 98.9% (IID) / 98.2% (non-IID); under poisoning ~95–99.1% accuracy, ASR ≈0.008–0.074; robust to ≤30% malicious clients. |
| Alsaleh, Menai, Al-Ahmadi (2025)[3]: A Heterogeneity-Aware Semi-Decentralized Model for a Lightweight IDS for IoT Networks Based on FL and BiLSTM | Semi-decentralized FL: client clustering + FedAvg; lightweight BiLSTM for efficiency | Train/val/test: CICIoT2023; Cross-dataset tests: BoT-IoT, WUSTL-IIoT-2021, Edge-IIoTset | CICIoT2023 99.43% (binary) / 98.55% (34-class); Edge-IIoTset 95.17%; WUSTL-IIoT-2021 99.96%; macro-F1 ≈0.67–0.94. |
| Gutti, Thumula & Balbudhe (2025)[4]: Federated Learning for Distributed IoT Security: A Privacy-Preserving Approach to Intrusion Detection | HADA-FL: adaptive aggregation using SHAP stability + DP budgets; robust to attacks | TabularIoTAttack-2024, Edge-IIoTset | 85–89% nominal; 66–73% under FGSM/PGD/Label-Flip |
| Fatema et al.(2025)[5]: Federated XAI IDS (Future Internet 2025) | ANN + FedAvg with SHAP for explainability | CICIoT2023 | 88.4% train / 88.2% test. |
| Hamad et al. (2025)[10]: Systematic Analysis of Federated Learning Approaches for Intrusion Detection in the IoT Environment | Systematic review of 43 FL-IDS studies (2020–2024). | Multiple IoT/general datasets | No model executed; best reported accuracies in surveyed works up to ≈99.6% (with large variance across datasets). |

In FL, individual devices (or clients) train local models using their private data and then share only the model updates with a central server. The server aggregates these updates to produce a global model, ensuring that raw data never leaves the local devices. This decentralized approach is particularly beneficial for IoT networks, where transmitting large volumes of sensitive data to a central server can be both impractical and insecure. Recent studies have applied FL to various cybersecurity applications, such as malware detection, spam filtering, and anomaly detection in network traffic and intrusion detection. These works demonstrate that FL can achieve performance comparable to centralized training while significantly reducing the risk of data breaches.

### *D. SHAP (SHapley Additive Explanations)*

Shapley Value is a concept from game theory, specifically referring to cooperative games. It is defined as the fair contribution of each member to the total value generated by a coalition. Mathematically, this value is computed by averaging the marginal contribution of each member across all possible sub-coalitions.

Lundberg and Lee introduced an approach called SHAP in 2017 [11], building upon the Shapley value concept. To map the Shapley value to ML problems, the features of the input vector are treated as the members of the coalition, and the model’s output is treated as the coalition's value.

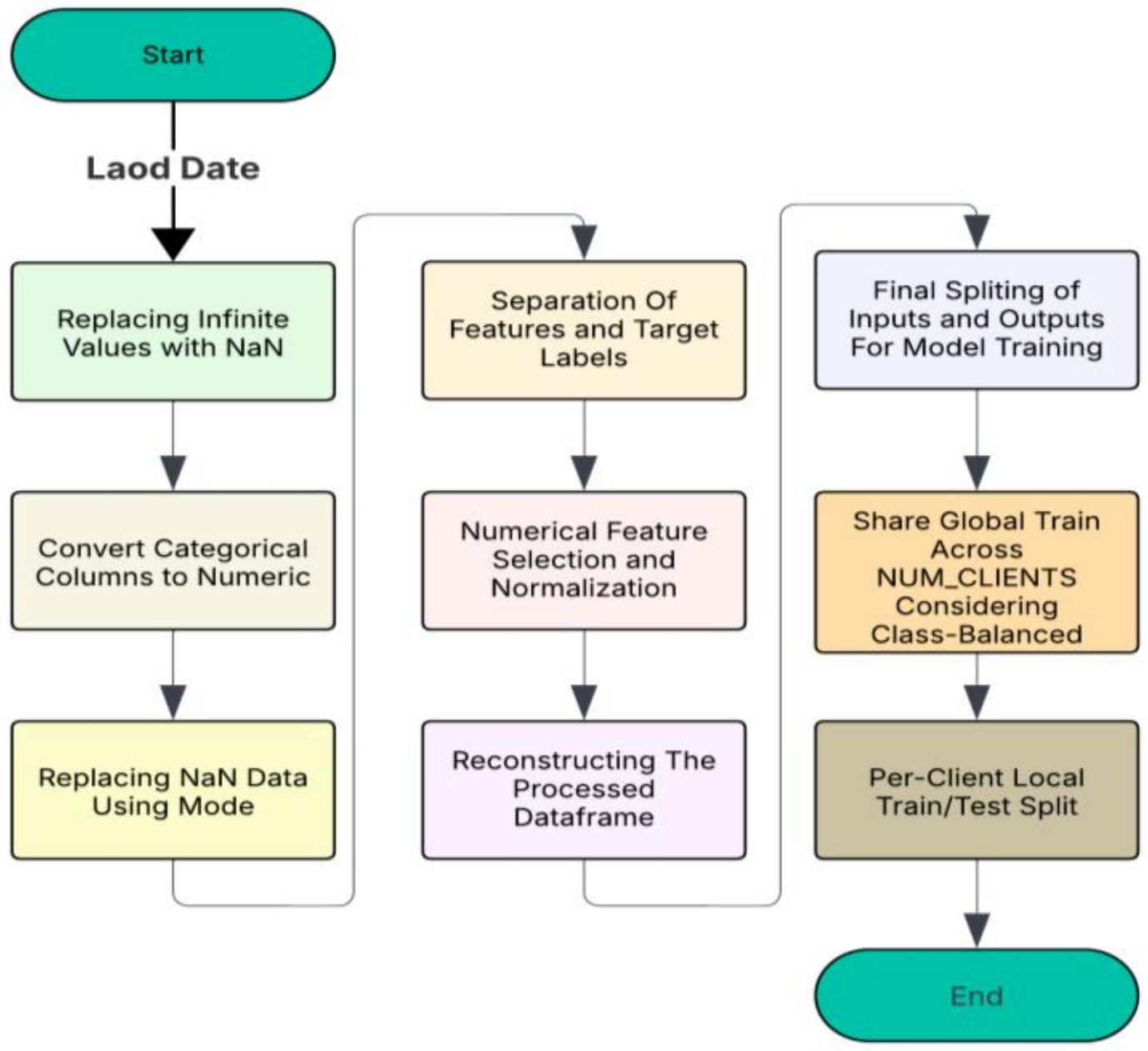

*Fig. 1. Data Processing Framework for XAI FL-IDS*

As the name suggests, this approach has an additive procedure, where the system adds to a baseline value based on the values of each feature in a sample, until it reaches the sample's predicted output. The values that are added represent the contribution of each feature to the final prediction result.

If we consider x as the model inputs and f(x) as the model's output, x′ is the simplified binary vector of x, indicating whether each feature is included in the model or omitted. G(x′) is the approximation of the original model f(x), which is defined by the following additive form:

$$G(x') = \varphi_0 + \sum_{i=1}^{N} \varphi_i \, x'_i \quad (1)$$

Where:

$\varphi_0$: Represents the baseline output of the model—specifically, the mean output when no features are considered.

$\varphi_i$: Denotes the contribution (or attribution) of feature to the model's output; this is the Shapley value for that specific feature.

Interpretability (Explainable AI - XAI) was integrated into the design from the outset. On each client, following evaluation, SHAP values are computed using the Tree-Explainer and saved along with detailed metadata, including the prediction ID, true/predicted label, output probability, client ID, and round number. On the server, after the global model is selected, SHAP is generated and archived for the entire global test set. This two-level explainability architecture enables the dynamic tracking of feature importance across two scales: *local* heterogeneity and *global* stability. Consequently, this allows researchers to observe shifts in the role of features both throughout the training process and across different clients.

### E. XAI FL-IDS

The system proceeds as follows in fig.2 : If the adequate number of epochs is obtained, after data processing, the system generates output from the log. Otherwise, each client trains the model sent by the server using its local data. Subsequently, clients compute performance metrics and SHAP values. If clients fail to generate any output, the current learning round is halted, and the event is saved to the log. The server then compares all client outputs and selects the best model for broadcasting in the next round. Finally, global metrics and SHAP values are calculated.

As Fig.3 suggests, each client, when parameters are received from the server, the model is trained through them. If no parameters are available, the client trains the model with a None parameter set. In the next step, class balancing is performed between the positive and negative data classes, and the model is trained. Finally, set of metrics (such as accuracy, F1-score, ROC-AUC) and uses Tree-Explainer to extract SHAP values for the entire local test set. The outputs, along with the client ID, are saved in per-round files, and any failure events are meticulously documented in the local log after that the model is sent back to the server.

Upon receiving the global model—if one exists—each client executes low-cost continuous boosting: training resumes precisely from that Booster object with num_boostround equal 1. If no parameters were received, the model is initialized from scratch. To mitigate local class imbalance, class weights are adaptively adjusted using scale_pos_weight = #neg / #pos. This assigns greater importance to the minority class without disrupting the stability of convergence.

In Fig.4, the server first receives parameters from clients. If no valid results are received, the system bypasses the current round, and the event is saved to the log. However, if valid results are received, the server first identifies the best model among clients based on the hierarchical scoring and broadcasts it to all clients for the next round. It then calculates the global SHAP value and aggregates clients' metrics. Finally, if there are remaining rounds to be executed, it proceeds to the next round; otherwise, all events—including rounds skipped due to a lack of valid output—are saved as structured data in the global log to facilitate scientific auditing and debugging.

The XGBoost Booster object is exchanged between the server and clients as a byte-serialized uint8 array [5]. This process ensures reproducibility while maintaining the confidentiality of the raw data. For global interpretability, the server computes and archives the global SHAP values using the selected model. The execution layer is implemented atop Ray, utilizing a minimal quota of one CPU per worker and no GPU acceleration.

## IV. PERFORMANCE EVALUATION

### A. Experimental Environment

The research experiment was conducted in the Google Colab environment (runtime: CPU), equipped with an Intel(R) Xeon(R) CPU (2.20 GHz) processor and 13.61 GB of main memory. The programming environment used was Jupyter Notebook, integrated with Python 3, within Colab as the framework to run the experiment. The following section discusses the outputs.

### B. Evaluation Metrics and Result

Given that the proposed approach is in the domain of Artificial Intelligence, the system evaluation metrics include Precision, Accuracy, and Recall, along with the F1-score.

Confusion matrix: The Confusion Matrix is an evaluation parameter used for classification type Artificial Intelligence models. It is completed by comparing the true labels of the samples against the system's predictions on those samples. This table consists of the following components: TP—true positive, TN—true negative, FP—false positive, and FN—false negative [5].

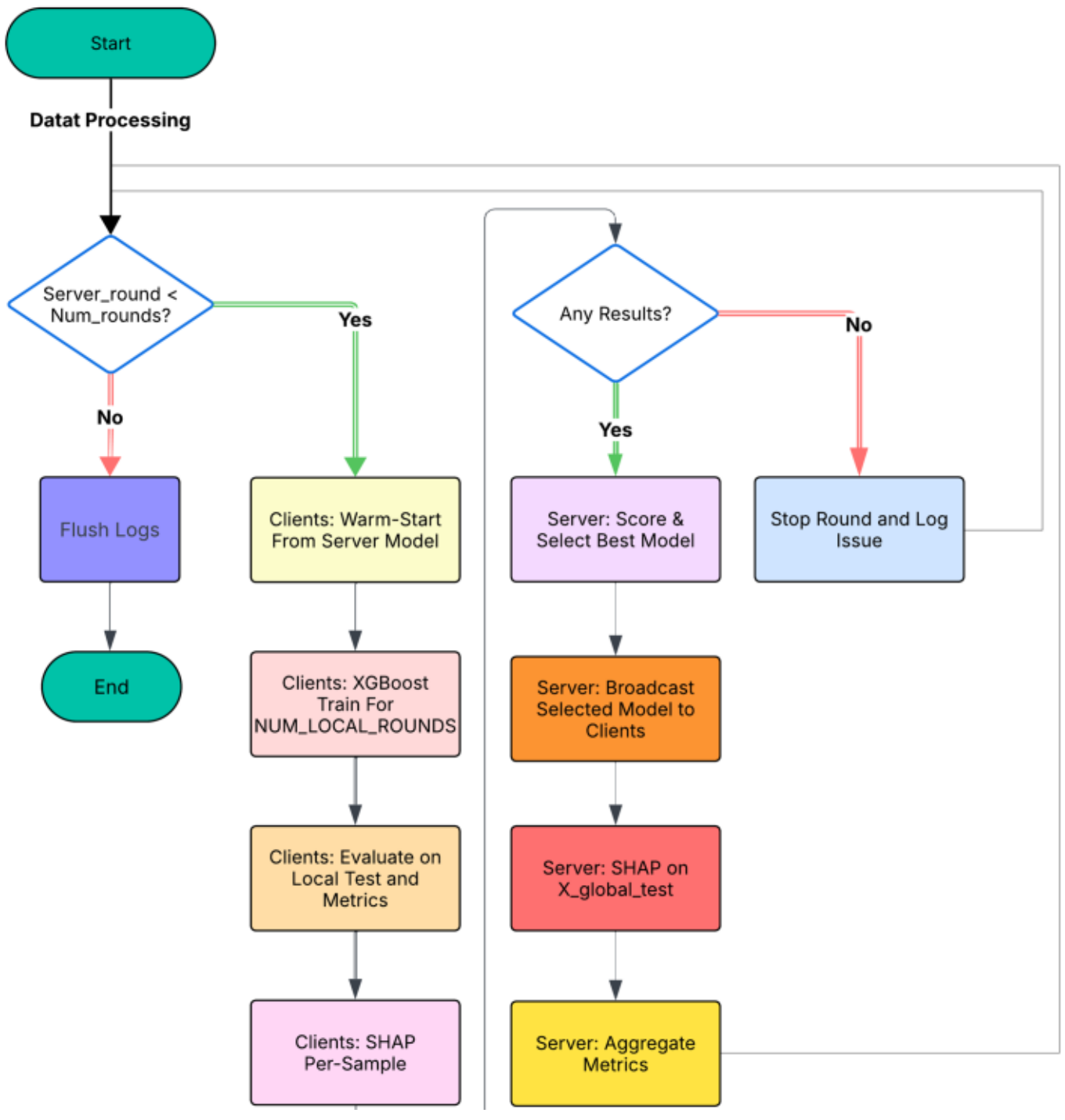

*Figure 2 : XAI FL-IDS Framework*

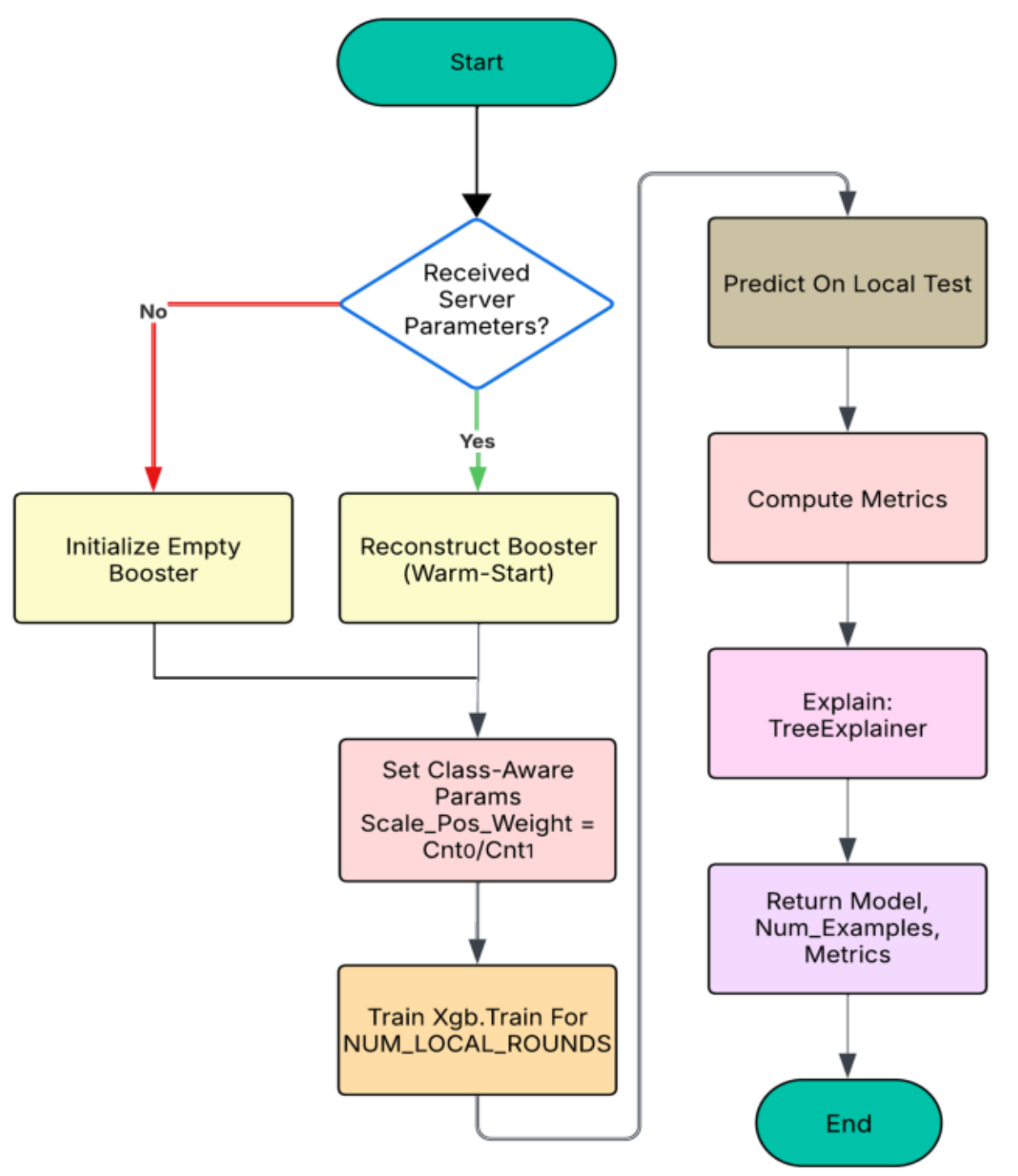

*Fig. 3. Client Side of XAI FL-IDS*

$$\text{Confusion Matrix} = \begin{bmatrix} TP & FN \\ FP & TN \end{bmatrix} \tag{2}$$

Accuracy: The metric that reflects how often a ML method correctly predicts an outcome is called accuracy. Accuracy is assessed by dividing the number of correct predictions by the total number of predictions made [5].

$$\text{ACCURACY} = \frac{TP + TN}{FP + TP + FN + TN} \tag{3}$$

Recall: Recall is the ability to identify all relevant instances within the dataset. It is defined as the ratio of true positives to the sum of true positives and false negatives in the field of mathematics [5].

$$\text{RECALL} = \frac{TP}{FN + TP} \tag{4}$$

Precision: Precision is the metric that measures the effectiveness of a ML approach. It indicates the accuracy of the algorithm's positive predictions. It is the ratio of true positives to the total number of positive predictions [5].

$$\text{PRECISION} = \frac{TP}{FP + TP} \tag{5}$$

F1-score: The F1-score is the balanced average of accuracy and recall. It combines accuracy and recall into a single metric to improve the understanding of the effectiveness of the proposed framework [5].

$$\text{F1-SCORE} = \frac{2 \times RECALL \times PRECISION}{RECALL + PRECISION} \tag{6}$$

Log-Loss : Logarithmic loss (or log loss) is a mathematical function that measures the discrepancy between the predicted probability distribution produced by a computational model and the actual categorical outcomes. The log loss function assesses how well the model's probability-based predictions align with the true class labels, imposing penalties on incorrect predictions in proportion to the model's confidence. Formally, for a dataset with N samples and M possible classes, the log loss is defined as:

$$\text{Log loss} = -\frac{1}{N} \sum_{i=1}^{N} \sum_{j=1}^{M} y_{ij} \cdot \log p_{ij} \tag{7}$$

Where:

$y_{ij}$ : {0,1} denotes the true class label (equal to 1 if sample i belongs to class j, and 0 otherwise)

$p_{ij}$: represents the predicted probability that sample i belongs to class j.

The established baseline solution achieved an accuracy of 99.7% on a comparable dataset. As illustrated in Fig.5, the proposed methodology in this article not only matched this high degree of precision but also, in certain instances, reached a perfect 100% accuracy. Furthermore, the results across all other metrics consistently demonstrate the superior efficiency and effectiveness of the XAI FL-IDS system under various conditions.

Consistent with the earlier explanation of Log Loss, Fig.6 illustrates the value of this metric in the proposed approach. A lower Log Loss value indicates superior system performance, signifying that the model assigns a higher probability to the correct outcome. Conversely, the Log Loss value increases significantly when the prediction is incorrect. This trend further validates the efficacy of the proposed XAI FL-IDS architecture.

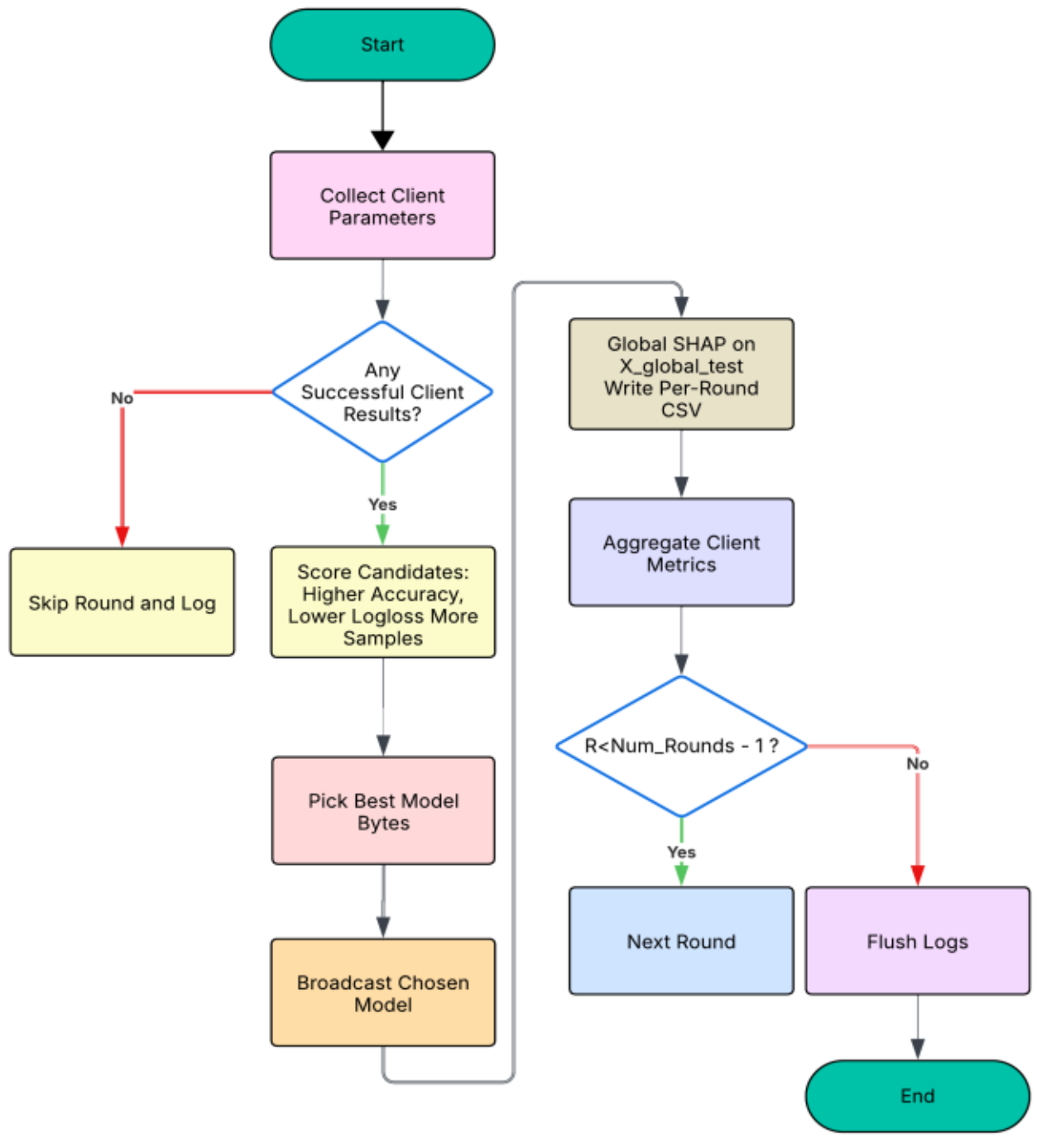

*Fig. 3. Server Side of XAI FL-IDS*

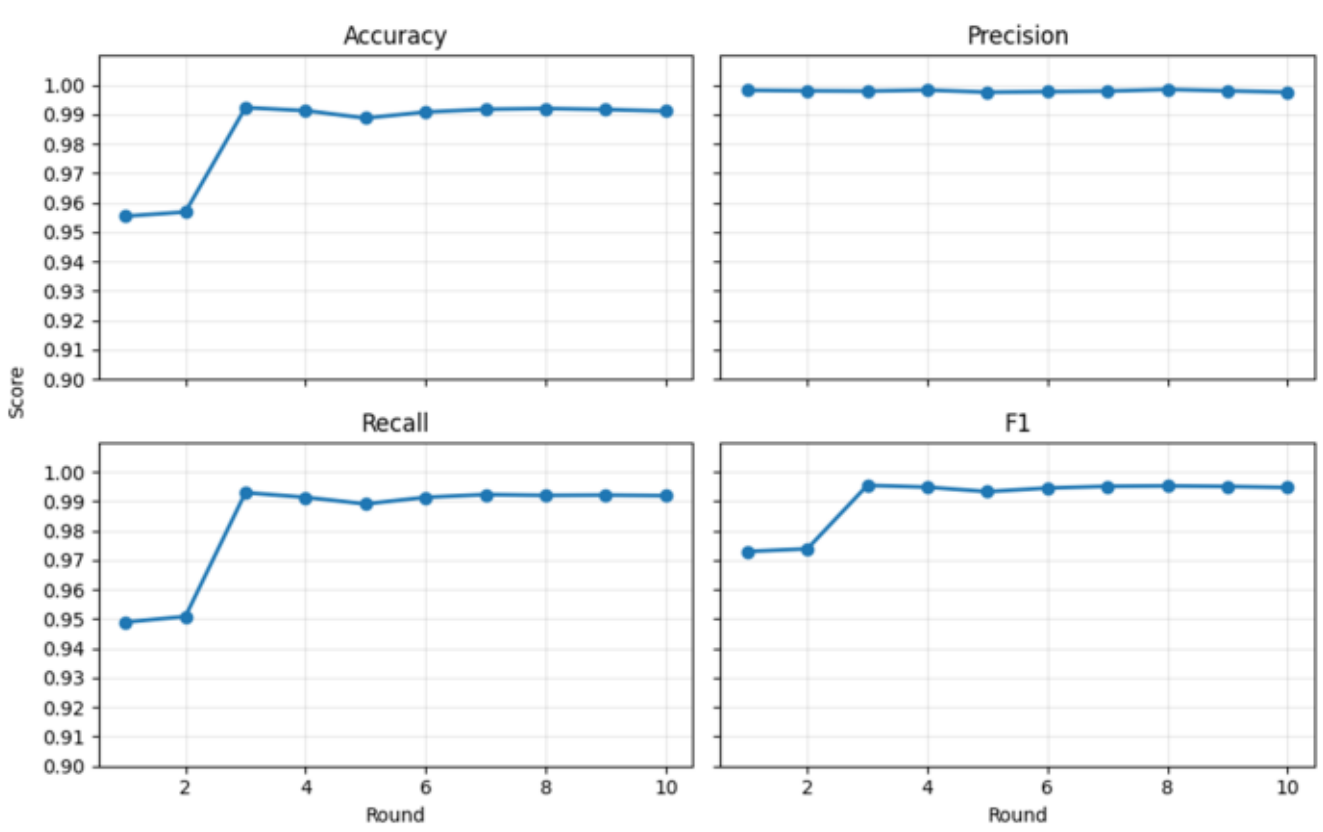

*Fig. 4. Diagram of XAI FL-IDS Evaluation Metrics*

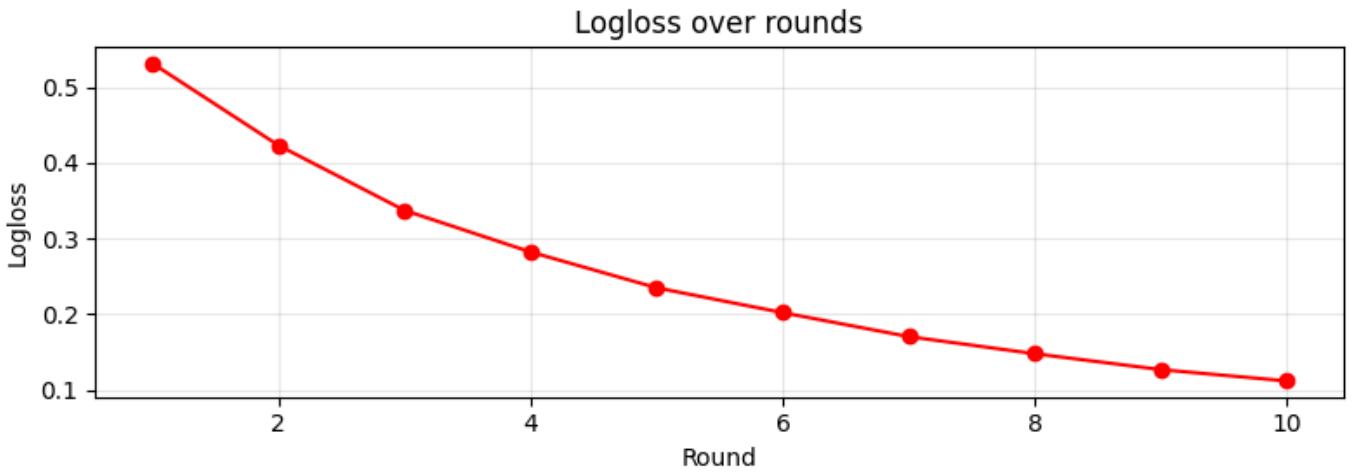

*Fig. 5. Diagram of Log Loss Metric of XAI FL-IDS*

By modifying the baseline method and replacing traditional averaging with the best-model selection policy, the training time has been significantly improved. As demonstrated in the Fig.7 and Fig.8, each client experiences a notably lower average training time with the new approach. Furthermore, each client actively prevents resource wastage by meticulously controlling and limiting superfluous learning rounds.

The final section details the outputs obtained from SHAP. As illustrated in the tables, one column indicates the true label of the data, and the adjacent column shows the system's prediction, with a corresponding SHAP Value calculated for each decision.

Table 2 displays the output from Client 5's first round, where the decision boundary for sample detection is 0.5. However, this value progresses over time, indicating an improvement in detection accuracy. Table 3 corresponding to Client 5's ninth round, shows that the decision boundary has increased to approximately 0.85, demonstrating the model's enhanced confidence and performance over the course of training.

## V. Conclusion

An Intrusion Detection System (IDS) is vital in cybersecurity, detecting unauthorized activity across networks. With attacks on network layers increasing, stronger IDSs are needed. Yet most IDSs rely on centralized detection, forcing IoT nodes to ship data to a server—adding overhead and offering no privacy guarantees. Moreover, conventional models focus solely on flagging attacks, without explaining how individual features influence those decisions.

This research aims to address these dual limitations by first proposing a solution for privacy preservation and then adding explainability to the new system. We introduce an innovative framework called XAI FL-IDS, which integrates FL with XAI.

The XAI FL-IDS system eliminates concerns over data transfer because each node trains its data locally and only sends the necessary update parameters to the server. Additionally, all detections, both at the local node and central server levels, are scrutinized using SHAP, providing detailed insight into the decision-making process. This system consists of a central server and 10 clients and utilizes the Edge-IIoTset dataset, which is distributed among all clients with careful attention paid to class balancing. On each client, the XGBoost model is executed on local data. After receiving client parameters, the central model selects the single best model to broadcast to all clients for the next round.

The proposed method demonstrates robust efficiency and strong performance in intrusion detection, achieving an accuracy of over 99% and, at times, reaching 100%. The SHAP outputs confirm that the model's understanding of features improves over time, offering increasingly detailed explanations of feature importance.

FL ensures the confidentiality of network information on each local node, enhancing transparency and making the model both reliable and interpretable. The XAI FL-IDS effectively tackles privacy concerns and feature interpretability challenges within modern IDS frameworks, advancing secure, interpretable, and decentralized IDS.

The continuous advancement of technology means the efforts to secure IoT systems must also persist. In the development pathway for the XAI FL-IDS framework, future work can explore utilizing diverse feature selection methods to further optimize the model, as well as evaluating its performance across a wider variety of datasets. Furthermore, given the low computational capacity of IoT nodes, adopting lighter-weight models would enable the current system to evolve into a more resource-efficient architecture, directly addressing critical device-level issues such as improving energy and memory efficiency on IoT devices.

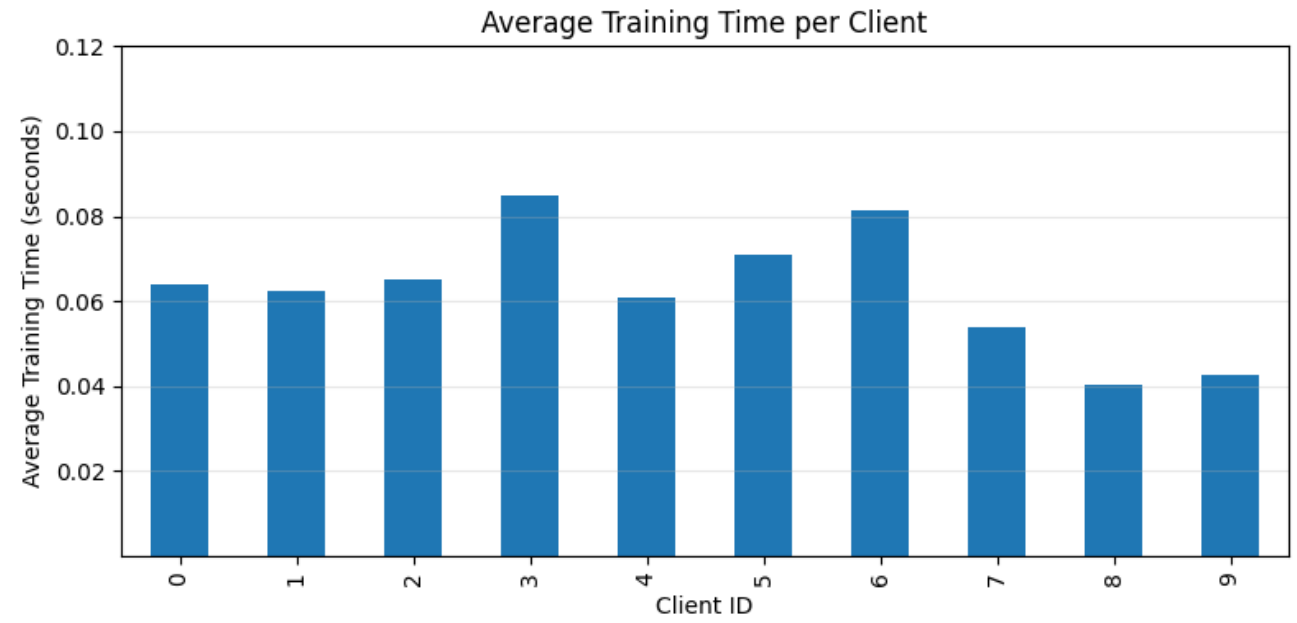

*Fig. 6. Average Training Time Per Client of XAI FL-IDS*

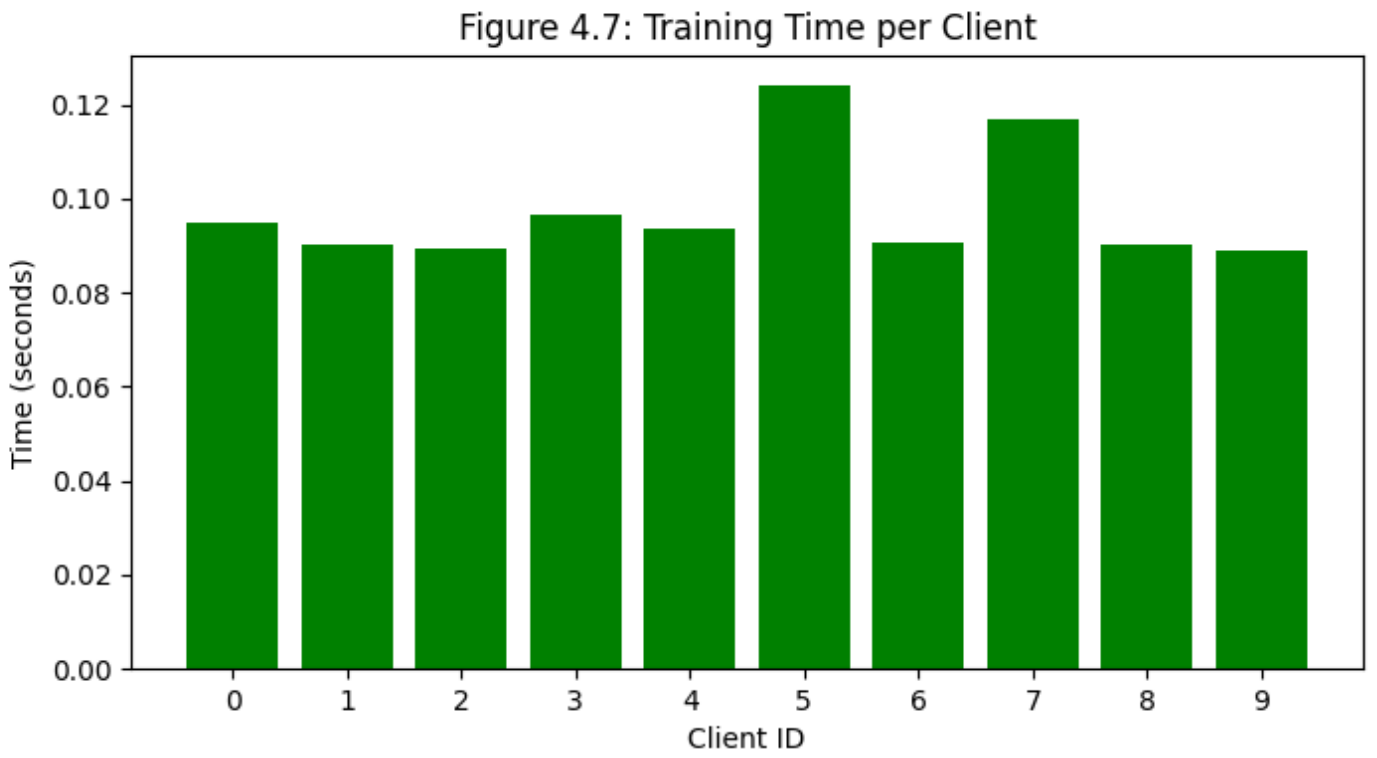

*Fig. 7. Average Training Time Per Client of Base Method*

*Table. 2. Detection Explanation in Round 1 of Client 5*

| PREDICTION_ID | TRUE_LABEL | PREB_LABEL | PREB_PROBABILITY |
|---|---|---|---|
| 0 | 1 | 1 | 0.5936770151 |
| 1 | 1 | 1 | 0.5979139805 |
| 2 | 0 | 0 | 0.4041649997 |
| 3 | 1 | 1 | 0.5963770101 |
| 4 | 1 | 1 | 0.5926724078 |
| 5 | 1 | 1 | 0. 5936770254 |
| 6 | 1 | 1 | 0.5968382955 |

*Table. 3. Detection Explanation in Round 9 of Client 5*

| PREDICTION_ID | TRUE_LABEL | PREB_LABEL | PREB_PROBABILITY |
|---|---|---|---|
| 0 | 1 | 1 | 0.901085794 |
| 1 | 1 | 1 | 0.8985581398 |
| 2 | 0 | 0 | 0.7687667012 |
| 3 | 1 | 1 | 0.8876637816 |
| 4 | 1 | 1 | 0.9054922462 |
| 5 | 1 | 1 | 0. 9333615303 |
| 6 | 1 | 1 | 0. 8775247935 |